\documentclass[aps,superscriptaddress,nofootinbib,notitlepage]{revtex4-1}
\usepackage{epsfig}
\usepackage{amsmath}
\newcommand{\ben}{\begin{eqnarray}}
\newcommand{\een}{\end{eqnarray}}
\newcommand{\nnu}{\nonumber\\}
\newcommand{\bef}{\begin{figure}[hbt]\centering}
\newcommand{\eef}{\end{figure}}

\begin{document}

\title{Nuclear modification of vector boson production in proton-lead collisions at the LHC}

\author{Zhong-Bo Kang}
\email{zkang@lanl.gov}
\affiliation{Theoretical Division,
                   Los Alamos National Laboratory,
                   Los Alamos, NM 87545, USA}
                   
\author{Jian-Wei Qiu}
\email{jqiu@bnl.gov}                   
\affiliation{Physics Department, 
                   Brookhaven National Laboratory, 
                   Upton, NY 11973, USA}
\affiliation{C.N. Yang Institute for Theoretical Physics, 
                   Stony Brook University, 
                   Stony Brook, NY 11794, USA}

\date{\today}

\begin{abstract}
In anticipating the upcoming proton-lead run at the LHC in the near future, we present predictions for the nuclear modification factor of transverse momentum spectrum of $Z^0$ production and transverse momentum broadening of vector boson ($J/\psi$, $\Upsilon$, $W/Z^0$) production in proton-lead collisions at $\sqrt{s}=5$ TeV, respectively.  We find that the measurement of nuclear modification factor of $Z^0$ production provides a clean and unambiguous test of the nuclear anti-shadowing proposed in the recent EPS09.  In addition, the dramatic difference in transverse momentum broadening between the heavy quarkonium and $W/Z^0$ production could be a signature prediction of QCD parton multiple scattering, and it provides valuable informations on color neutralization of a produced heavy quark pair when it transmutes into a physical quarkonium.
\end{abstract}

\maketitle

\section{Introduction}
Extensive studies of ultra-relativistic nucleus-nucleus collisions carried out by the experiments at the Relativistic Heavy Ion Collider (RHIC) at BNL and the Large Hadron Collider (LHC) at CERN have demonstrated that the hot and dense matter of quarks and gluons, the quark-gluon plasma (QGP) was formed in the interactions of heavy nuclei at high energies \cite{RHIC,LHC}. In order to extract precise information on QGP properties, one needs to calibrate hard probes by producing them in better-controlled cold nuclear matter.  The upcoming proton-lead (p+Pb) collisions at the LHC should accomplish precisely that, thereby provide crucial benchmarks for understanding the characteristic of QGP produced in heavy ion collisions at the LHC \cite{Salgado:2011wc}. 

In anticipating this upcoming p+Pb run at the LHC in the near future, we propose in this paper two measurements to better derive nuclear parton distribution functions (PDFs) and to study the dynamics of parton multiple scattering in the cold nuclear matter. In the first part of our paper, we will study the nuclear modification of transverse momentum spectrum for neutral heavy boson $Z^0$ production in p+Pb collisions comparing with that in p+p collisions, and demonstrate that the modification serves as a direct measurement of nuclear modification of gluon distribution. Then, we study the transverse momentum broadening of vector boson ($J/\psi$, $\Upsilon$, $W/Z^0$) production in p+Pb collisions to explore the dynamical consequences of parton multiple scattering in QCD and the formation of heavy quarkonia from produced heavy quark pairs.

Production of $Z^0$ bosons, particularly the full transverse momentum spectrum in p+p collisions, has been a test ground for short-distance dynamics of QCD and an ideal observable for perturbative QCD treatment of hard probes.  This is because of the nature of two observed hard scales, the $Z^0$-boson mass $M_Z$ and its transverse momentum $p_T$, and our ability to measure a very wide range of $p_T$.  When $p_T\sim M_Z$, $Z^0$-boson production is a very clean probe of perturbative QCD at a distance scale as small as an attometer.  On the other hand, when the two scales are very different, the $p_T$-distribution of $Z^0$ boson production could be a perfect laboratory to study the richness of QCD dynamics covered by these two scales as well as the influence of non-perturbative physics.  At low transverse momentum $p_T\ll M_Z$,  the $p_T$-distribution, calculated in the conventional fixed-order perturbation theory, receives a large logarithm, $\ln(M_Z^2/p_T^2)$. One could get as many as two powers of the logarithm for every power of coupling constant $\alpha_s$ when the calculation goes beyond the leading order. Therefore, at a sufficiently small $p_T$, convergence of the conventional perturbative expansion is impaired, and these large logarithms must be resummed. A resummation formalism, known as Collins-Soper-Sterman (CSS) formalism, was proposed, developed and improved over the years \cite{CSS-resum,Ellis:1997ii,Qiu:2000ga,Qiu:2000hf,ResBos,Gavin:2010az}, and has become one of many successful examples in perturbative QCD toolkit. 

The study of nuclear modification of $Z^0$ production has finally become available at the LHC \cite{Chatrchyan:2011wt,Chatrchyan:2011ua,Aad:2011gj,Aad:2012ew,Aad:2010aa,delaCruz:2012ru}. Nuclear modification of the inclusive $Z^0$ production and its implication on nuclear PDFs was studied in Ref.~\cite{Paukkunen:2010qg}.   There are also studies on nuclear modification for low mass virtual photon production but at high transverse momentum  \cite{Kang:2008wv}. In this paper, we will use the CSS resummation formalism to study nuclear modification to the $p_T$-distribution of $Z^0$ production in p+A collisions at the LHC. In proton-nucleus collisions, there are two major nuclear effects that one needs to consider carefully. One is the power corrections coming from the initial-state multiple parton scattering, which could be enhanced by the nuclear size and thus potentially could be a large effect. The other is from nuclear PDFs within the factorized leading-twist formalism \cite{Eskola:2009uj,deFlorian:2003qf,deFlorian:2011fp,Hirai:2007sx}. We will show below that the nuclear size enhanced power corrections (or the high-twist effect) are small for $Z^0$ production at the LHC energies.  But, 
the nuclear modification of $Z^0$ production is directly tied to the nuclear modification of PDFs, and thus, provides very good constraints to the nuclear PDFs. 

One of the sensitive observables to the parton multiple scattering is the transverse momentum broadening \cite{Kang:2008us,Kang:2011bp,Xing:2012ii}. For colorless vector boson production ($W/Z^0$), the incoming (anti-)quark from the proton can undergo multiple scattering with the soft partons inside the nuclear matter before the hard collision to produce the vector boson. We call them ``initial-state parton multiple scattering'', which will lead to an accumulative change to the averaged transverse momentum square $\langle p_T^2\rangle$.  This phenomenon is often referred to as transverse momentum broadening. On the other hand, heavy quarkonium production in proton-nucleus collisions could have both initial-state and final-state multiple scatterings in nuclear medium since the heavy quarkonium is very unlikely to be produced at the moment when the heavy quark pair was produced \cite{Brodsky:1988xz,Kang:2011mg}.  The final-state multiple scattering between the produced heavy quark pair and the nuclear medium before the pair transmutes itself into a physical quarkonium is sensitive to the color, spin and other characteristics of the pair, as well as the hadronization mechanism of heavy quarkonia.  That is, the nucleus in p+Pb collisions could serve as a femtometer size detector or filter to gain much needed information on the formation of heavy quarkonia.  The additional final-state interaction for heavy quarkonium production could lead to more broadening in quarkonia's transverse momentum.  The amount of additional broadening is determined by QCD multiple scattering and the formation mechanism of the quarkonia.  The dramatic difference in the transverse momentum broadening between heavy quarkonium ($J/\psi$, $\Upsilon$) and the $W/Z^0$ production, as presented later in this paper, is a signature prediction of QCD calculation of multiple scattering and the role of color. 

The rest of our paper is organized as follows. In Sec.~II, we study the nuclear modification of $Z^0$ boson production in p+Pb collisions at the LHC. In Sec.~III, we present predictions for the transverse momentum broadening of $J/\psi$, $\Upsilon$, and $W/Z^0$ productions. We conclude our paper in Sec.~IV.

\section{nuclear modification factor for transverse momentum spectrum of $Z^0$ production}

In order to cover the full range of transverse momentum of $Z^0$ bosons, 
we use the CSS resummation formalism \cite{CSS-resum} to calculate the $Z^0$ production in p+p collisions at the LHC energy, 
\begin{eqnarray}
\frac{d\sigma_{A+B\rightarrow Z^0+X}}{dy\, dp_T^2} &=&
\frac{1}{(2\pi)^2}\int d^2b\, e^{i\vec{p}_T\cdot \vec{b}}\,
\tilde{W}(b,M_Z,x_A,x_B) 
+ Y(p_T,M_Z,x_A,x_B)\, ,
\label{css-gen}
\end{eqnarray}
where $\tilde{W}$ term gives dominant contribution when
$p_T\ll M_Z$, and the $Y$ term is perturbatively calculable and given in Ref.~\cite{Qiu:2000hf}.  The $Y$ term 
allows the formalism to have a smooth transition from the resummed low $p_T$ region 
to a region where $p_T\sim M_Z$ and the fixed order perturbative QCD calculations work well.  
In Eq.~(\ref{css-gen}), $x_A= e^y\, M_Z/\sqrt{s}$ and 
$x_B= e^{-y}\, M_Z/\sqrt{s}$ with the rapidity $y$ and collision
energy $\sqrt{s}$.  The $\tilde{W}$ in Eq.~(\ref{css-gen}) 
is given by \cite{Qiu:2000hf}
\begin{equation}
\tilde{W}(b,M_Z,x_A,x_B) = \left\{
\begin{array}{ll}
 \tilde{W}^{\rm Pert}(b,M_Z,x_A,x_B) & \quad \mbox{$b\leq b_{max}$} \\
 \tilde{W}^{\rm Pert}(b_{max},M_Z,x_A,x_B)\,
 \tilde{F}^{NP}(b,M_Z,x_A,x_B;b_{max})
                        & \quad \mbox{$b > b_{max}$}
\end{array} \right.
\label{qz-W-sol}
\end{equation}
where $b_{max} = 1/$(few GeV) is a parameter to specify the region in which
$\tilde{W}^{\rm Pert}$ is perturbatively valid, and $\tilde{F}^{NP}$ is a non-perturbative function 
that determines the large $b$ behavior of $\tilde{W}$ and is defined below.  
In Eq.~(\ref{qz-W-sol}), the $\tilde{W}^{\rm Pert}(b,M_Z,x_A,x_B)$ 
includes all powers of large perturbative logarithms resummed  
from $\ln(1/b^2)$ to $\ln(M_Z^2)$ and has the following form 
\cite{CSS-resum} 
\begin{equation}
\tilde{W}^{\rm Pert}(b,M_Z,x_A,x_B) = 
{\rm e}^{-S(b,M_Z)}\, \tilde{W}^{\rm Pert}(b,c/b,x_A,x_B)\, ,
\label{css-W-sol}
\end{equation}
where $c$ is a constant of order one \cite{CSS-resum,Qiu:2000hf}, and
$S(b,M_Z) = \int_{c^2/b^2}^{M_Z^2}\, 
  \frac{d{\mu}^2}{{\mu}^2} \left[
  \ln\left(\frac{M_Z^2}{{\mu}^2}\right) 
  A(\alpha_s({\mu})) + B(\alpha_s({\mu})) \right],
$
with perturbatively calculated coefficients $A(\alpha_s)$ and $B(\alpha_s)$ 
given in Ref.~\cite{Qiu:2000hf} and references therein.  In Eq.~(\ref{css-W-sol}), 
$\tilde{W}^{\rm Pert}(b,c/b,x_A,x_B)$ has no large logarithms and is given by
\begin{equation}
\tilde{W}^{\rm Pert}(b,\frac{c}{b},x_A,x_B) =\sigma_0 \sum_{i=q,\bar{q}}
f_{i/A}(x_A,\mu=\frac{c}{b})\, f_{\bar{i}/B}(x_B,\mu=\frac{c}{b})\, ,
\label{css-W-pert}
\end{equation}
where $\sigma_0$ is the lowest order partonic cross section for a quark and an antiquark to produce a $Z^0$ boson \cite{Qiu:2000hf}, and the functions $f_{i/A}$ (and $f_{\bar{i}/B}$) are the modified parton
distributions given by  \cite{CSS-resum} 
\begin{equation}
f_{i/A}(x_A,\mu) = \sum_a 
  \int_{x_A}^1\frac{d\xi}{\xi}\, 
  C_{i/a}(\frac{x_A}{\xi},\mu))\, \phi_{a/A}(\xi,\mu)
\label{mod-pdf}
\end{equation}
where $\sum_a$ runs over all quark flavors as well as gluon, 
$\phi_{a/A}(\xi,\mu)$ are the normal proton or effective nuclear PDFs 
with parton flavor $a=q, \bar{q}, g$, 
and $C_{i/a}=\sum_{n=0} C_{i/a}^{(n)} (\alpha_s/\pi)^n$
are perturbatively calculable coefficient functions for finding a
parton $i$ from a parton $a$, which are given in Ref.~\cite{Qiu:2000hf}.  

The non-perturbative function $\tilde{F}^{NP}$ in Eq.~(\ref{qz-W-sol}) has the following functional form, 
\begin{eqnarray}
F^{NP}(b,M_Z,x_A,x_B;b_{max}) 
& = & 
\exp\Bigg\{ -\ln\left(\frac{M_Z^2 b_{max}^2}{c^2}\right) \left[
    g_1 \left( (b^2)^\alpha - (b_{max}^2)^\alpha\right)
   +g_2 \left( b^2 - b_{max}^2\right) \right] 
\nonumber \\
&\ & {\hskip 0.8in}
   -\bar{g}_2 \left( b^2 - b_{max}^2\right) \Bigg\}\, .
\label{qz-fnp-m}
\end{eqnarray}
where the explicit $\ln(M_Z^2\,b_{max}^2/c^2)$ dependence was derived
by solving Collins-Soper equation \cite{CSS-resum}, the ${g}_2$ term 
is a result of adding a general power correction to the renormalization group equation,
and the $\bar{g}_2$ term represents the size of intrinsic transverse momentum of active partons 
\cite{Qiu:2000hf}. It is important to emphasize here that the actual size of $g_2$ signals the size of dynamical power corrections in our formalism \cite{Qiu:2000hf}. This will be important when we discuss the nuclear effects for $Z^0$ production in proton-nucleus collisions.

For a given choice of $b_{max}$ and $Z^0$ mass, 
coefficients of two terms proportional to $b^2$ in Eq.~(\ref{qz-fnp-m}) 
can be combined together as \cite{Zhang:2002yz},
\ben
G_2 = \ln\left(\frac{M_Z^2b_{max}^2}{c^2}\right)\, g_2 + \bar{g}_2\, ,
\een
which sums up both the dynamical and intrinsic power corrections. 
By requiring the first and second derivatives of $\tilde{W}$ to be continuous at $b=b_{max}$, the parameters: $\alpha$ and $g_1$ in Eq.~(\ref{qz-fnp-m}) can be uniquely fixed leaving only one parameter, $G_2$, sensitive to the power correction and other nonperturbative effects.  It was found \cite{Zhang:2002yz} that calculated $p_T$-distribution is insensitive to the choice of $b_{max}$ if it varies between 0.3~GeV$^{-1}$ and 0.7~GeV$^{-1}$.  Using $\bar{g}_2=0.25\pm 0.05$ GeV$^2$, and $g_2=0.01\pm 0.005$ GeV$^2$, and $b_{max}=0.5$~GeV$^{-1}$, effectively, $G_2^{pp} = 0.324$~GeV$^2$ \cite{Qiu:2000hf,Zhang:2002yz} in p+p collisions, theoretical predictions using Eq.~(\ref{css-gen}) are consistent with all data from Tevatron and the LHC \cite{Qiu:2000hf}, see Fig.~\ref{pp-compare} in which we see a very good agreement between the theoretical calculations and both the CDF \cite{Affolder:1999jh} and CMS \cite{Chatrchyan:2011wt} $Z^0$ experimental data at $\sqrt{s}=1.8$ TeV and $\sqrt{s}=7$ TeV, respectively.
\bef
\psfig{file=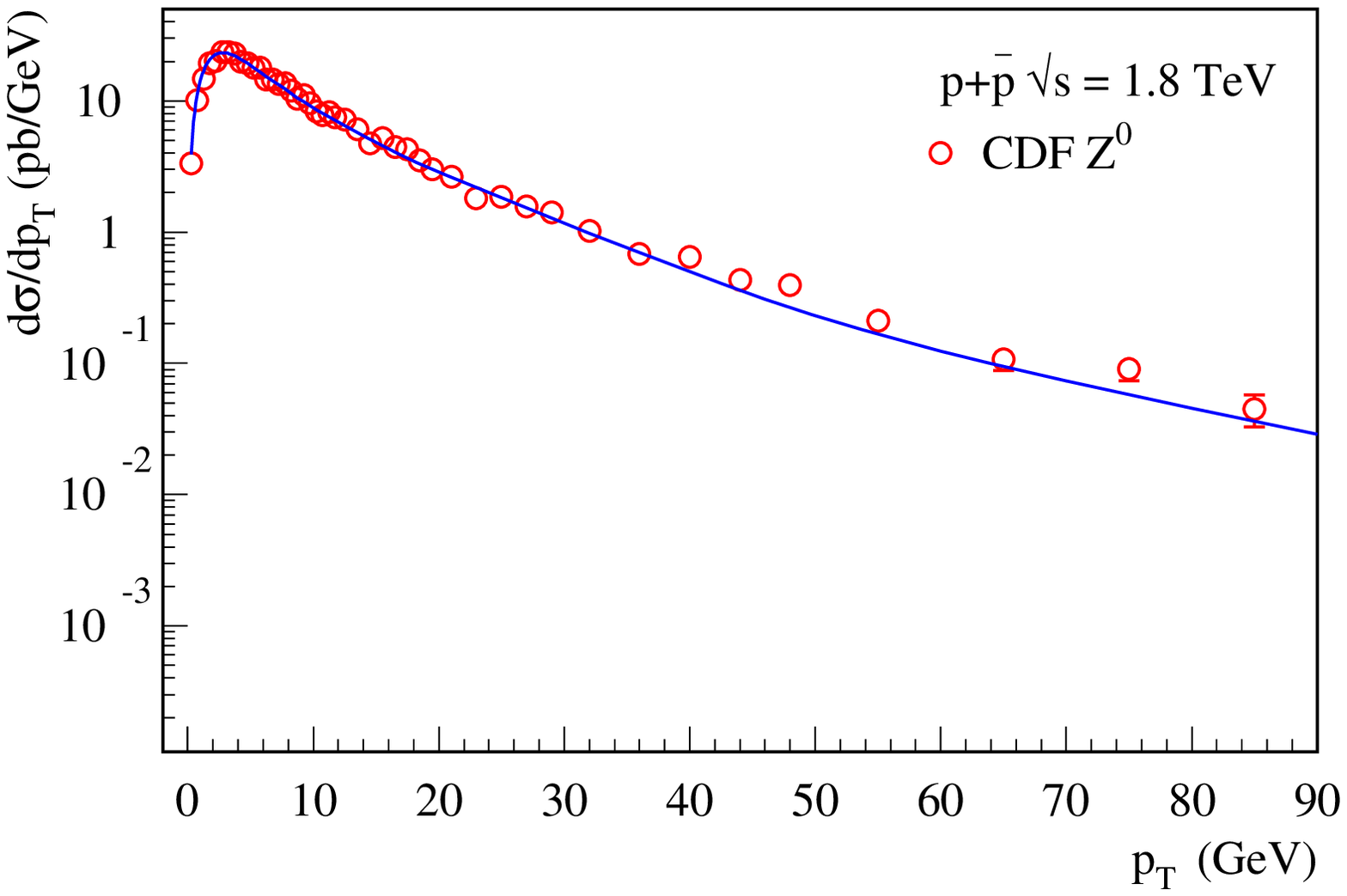, width=3in}
\hskip 0.3in
\psfig{file=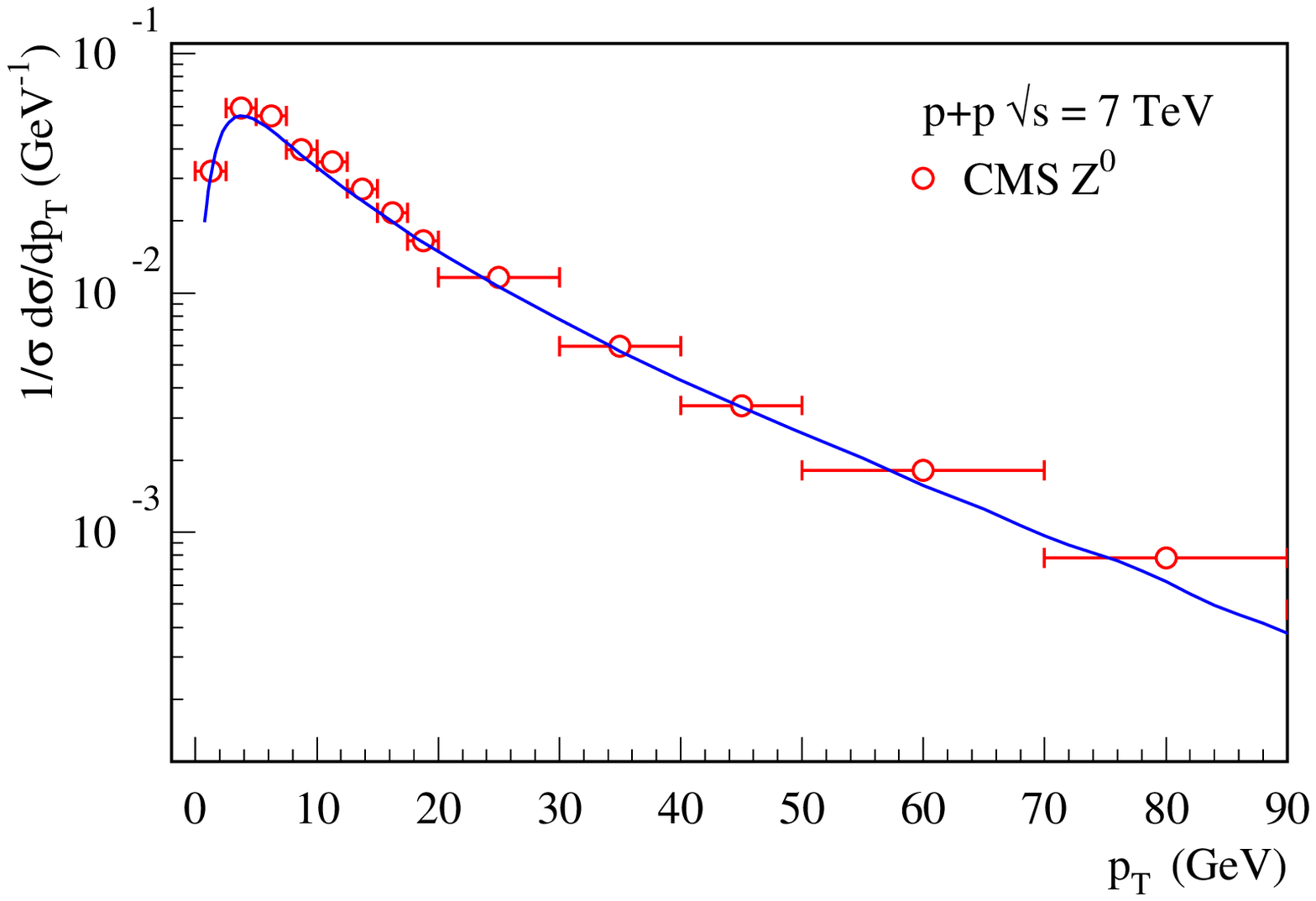, width=3in}
\caption{Theoretical calculations as in Eq.~\eqref{css-gen}  for $Z^0$ production in p+p (or p+$\bar{\rm p}$) collisions  are compared with experimental data from CDF \cite{Affolder:1999jh} and CMS \cite{Chatrchyan:2011wt} at $\sqrt{s}=1.8$ TeV and $\sqrt{s}=7$ TeV, respectively.}
\label{pp-compare}
\eef

For proton-nucleus collisions, we use EPS09 or nDS nuclear PDFs as parametrized in Refs.~\cite{Eskola:2009uj} and \cite{deFlorian:2003qf} to take care of the leading twist nuclear effect in PDFs. Although both parametrizations are based on the global fitting of experimental data, there is a major difference: EPS09 has an anti-shadowing region for gluon distribution inside a large nucleus, while nDS has no such region. We will see immediately the consequence of such a difference in the nuclear modification of $Z^0$ production. At the same time, there could be nuclear size ($\propto A^{1/3}$) enhanced multiple scattering effects. Such nuclear effects are usually manifest themselves as power corrections in the QCD factorization formalism \cite{Albacete:2013ei}. As we emphasize already that $g_2$ represents the size of dynamical power corrections in our resummation formalism, we could take into account these nuclear size enhanced multiple scattering effects by letting $g_2\to g_2 A^{1/3}$, following the method proposed in Ref.~\cite{Zhang:2002yz}. For $Z^0$ production, we have effectively, $G_2^{pP_b} = 0.689 {\rm ~GeV}^2$ in p+Pb collisions at LHC.

In Fig.~\ref{fig:z0}, we plot our predictions for $Z^0$ production at the LHC energy $\sqrt{s}=5$ TeV and mid-rapidity $y=0$ for the planned p+Pb run by evaluating the cross section with resummation in Eq.~(\ref{css-gen}) with CTEQ6M NLO parton distribution functions and the factorization scale $\mu=M_T/2=\sqrt{M_Z^2+p_T^2}/2$.  We evaluate the $Y$-term at NLO in $\alpha_s$ \cite{Qiu:2000hf}. The left panel is for EPS09 nuclear PDFs, while the right panel is for nDS nuclear PDFs. For both panels,
the upper plots are for the cross section of $Z^0$ production as a function of its transverse momentum.  The black dashed curve is the p+p baseline, and the red solid curve is for the minimum bias p+Pb collision.  The blue dotted curve is exactly the same as the red solid curves except that $g_2$ is not enhanced by $A^{1/3}$. That is, the nuclear size enhanced dynamical power corrections from multiple scattering is absent in blue dotted curve. Lower plots are for the nuclear modification factor $R_{pA}$ in min.bias p+Pb collisions. 
\bef
\psfig{file=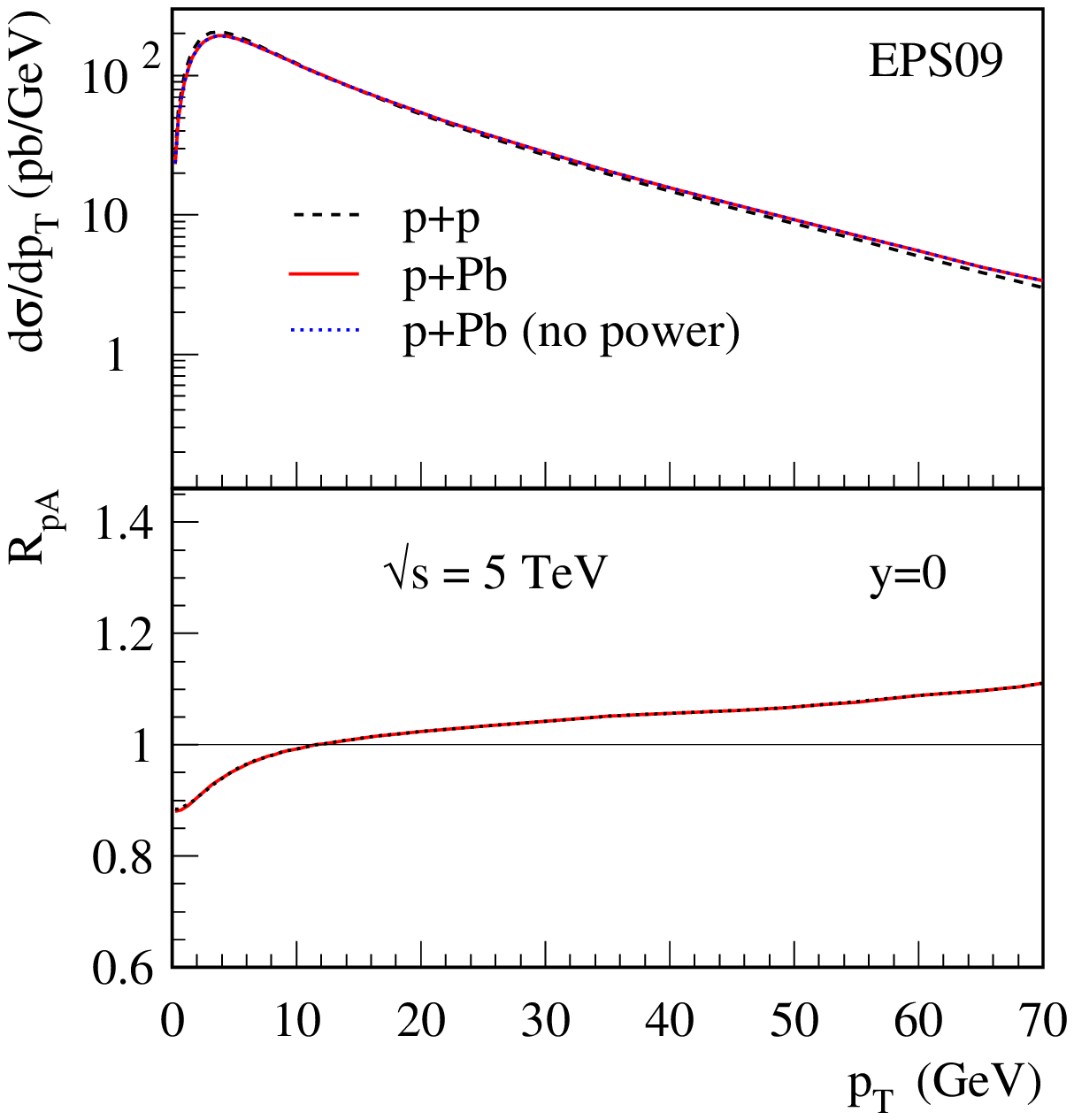, width=3in}
\hskip 0.3in
\psfig{file=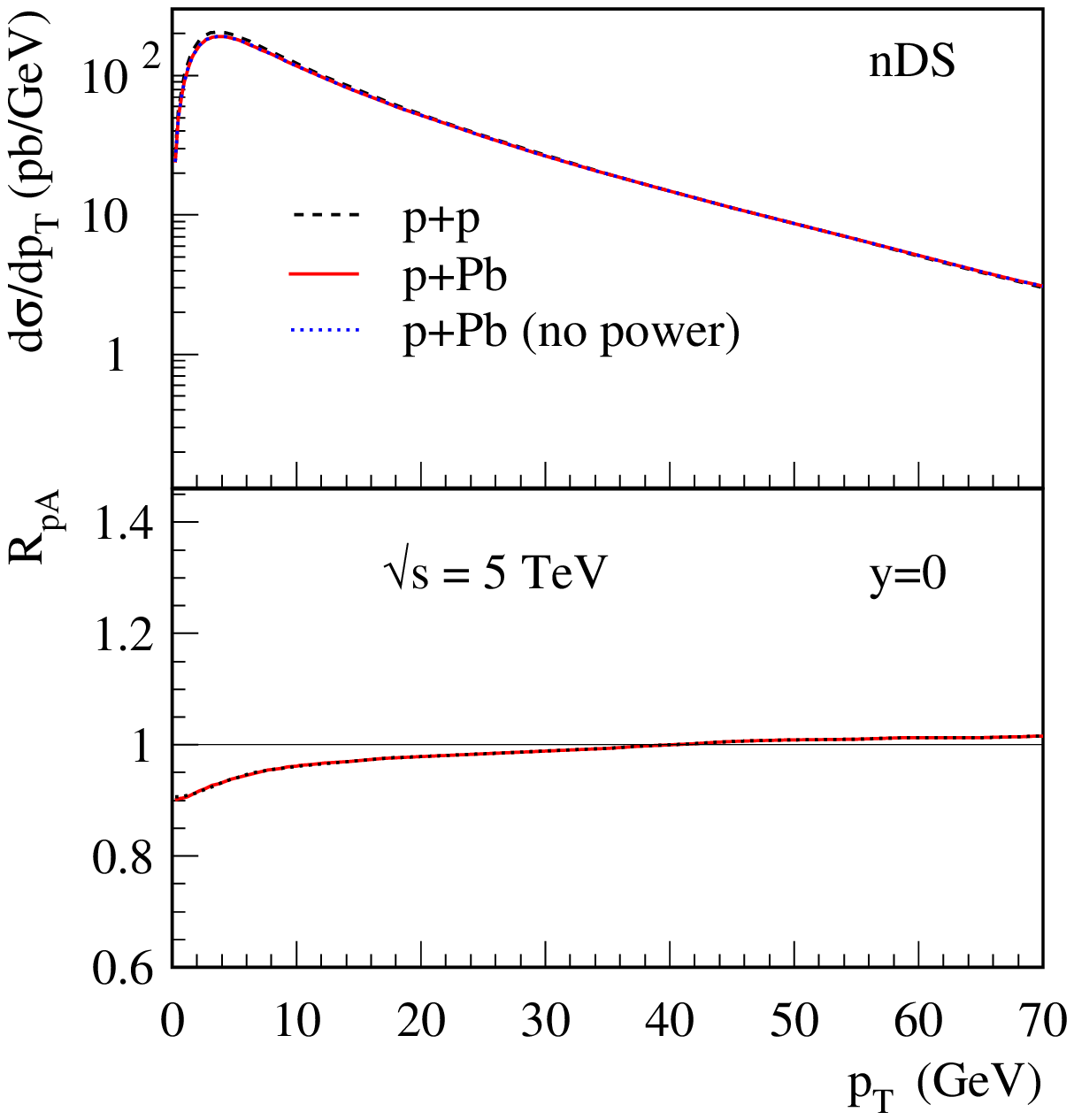, width=3in}
\caption{$Z^0$ boson production in both p+p and p+Pb collision at center of mass energy $\sqrt{s}=5$ TeV and rapidity $y=0$. The left panel is for EPS09 \cite{Eskola:2009uj}, while the right panel is for nDS \cite{deFlorian:2003qf}. For both panels, the upper plots are the $Z^0$ cross sections per nucleon as a function of transverse momentum $p_T$. The black dashed curve is the p+p baseline, and the red solid curve is for the minimum bias p+Pb collision. The blue dotted curve is exactly the same as the red solid curves except that $g_2$ is not enhanced by $A^{1/3}$. That is, the nuclear size enhanced dynamical power corrections from multiple scattering is absent in blue dotted curve.   Lower panel is for the nuclear modification factor, $R_{pA}$. }
\label{fig:z0}
\eef

From Fig.~\ref{fig:z0}, it is clear that the red solid curves are almost indistinguishable from the blue dotted curves.  That is, the power correction is indeed not important for $Z^0$ boson production at the LHC energy, and therefore, the $Z^0$ production in p+A collisions is an ideal process to probe nuclear modification of parton distribution functions, and to test the ``isospin'' effect at high energies. At the same time, we also find that the cross section of $Z^0$ production at $\sqrt{s}=5$ TeV and $y=0$ is dominated by the gluon initiated subprocesses when $p_T > 20$~GeV. This can be clearly seen in Fig.~\ref{gluon-contribution},  in which we plot the relative contributions (ratios) for $Z^0$ production in p+p collisions from the processes which involve gluon distribution (the red solid curve) and those which do not (the blue dashed curve). That is, $R_{pA}$ is an excellent observable to measure the nuclear modification of gluon distribution, which is effectively unknown. 
\bef
\psfig{file=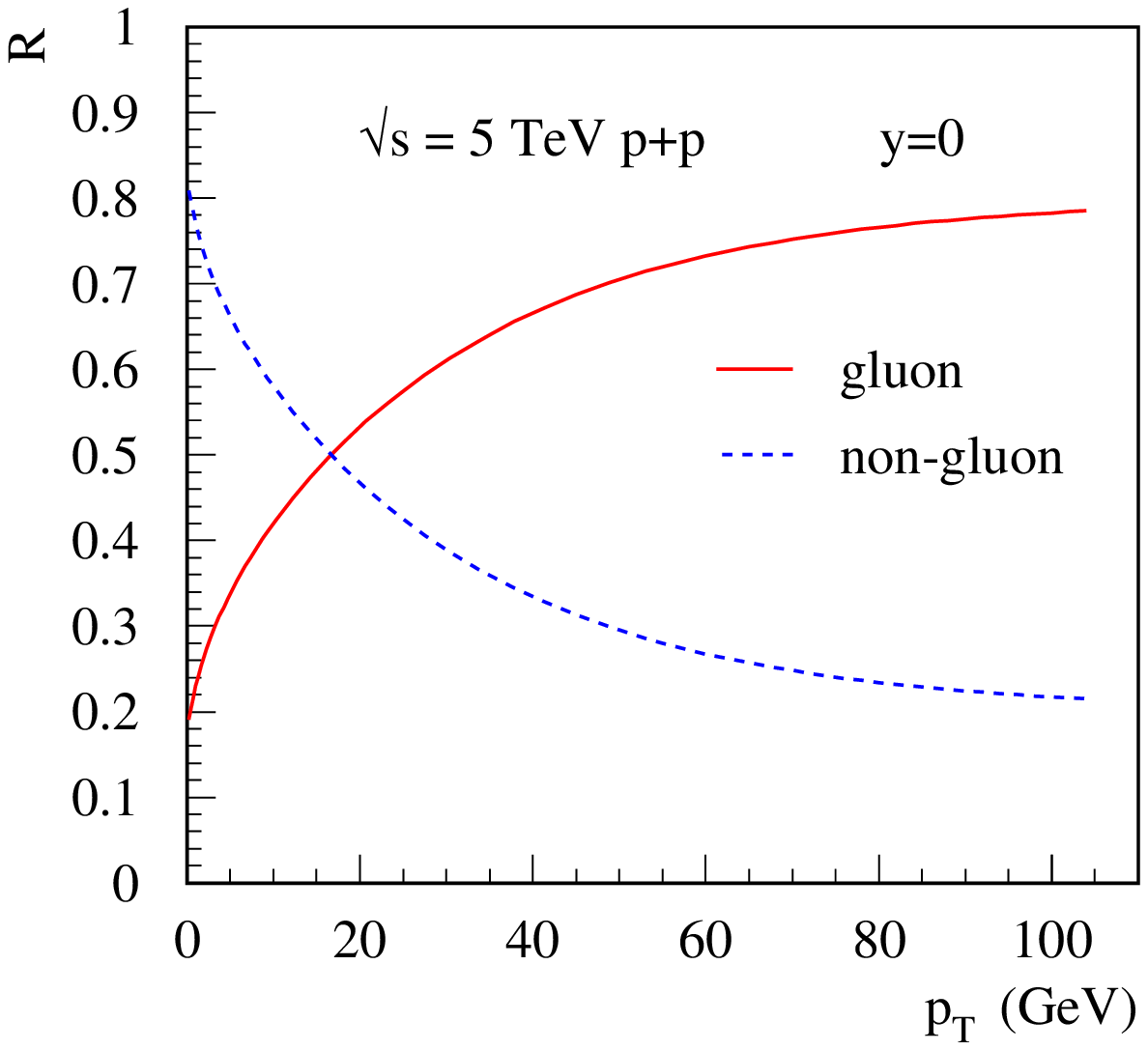, width=3in}
\caption{The fractional contributions to $Z^0$ production in p+p collisions at $\sqrt{s}=5$ TeV and rapidity $y=0$ from partonic subprocesses with at least one initial-state gluon (the red solid curve) and those not initiated by gluons (the blue dashed curve). }
\label{gluon-contribution}
\eef

Before we try to analyze and understand the nuclear modification factor $R_{pA}$ of $Z^0$ production in p+Pb collisions. Let us first take a look at the difference between EPS09 and nDS nuclear PDFs. In Fig.~\ref{fig:eps09}, we plot the nuclear modification ratio $R_i^A(x, \mu)$ at factorization scale $\mu=M_{Z}$ inside a lead nucleus for both valence $u$-quark and gluon distribution for EPS09 and nDS nuclear PDFs. Even though the valence $u$-quark distribution of both parametrizations is very similar, the gluon distribution is quite different: gluon distribution of EPS09 nuclear PDFs \cite{Eskola:2009uj} shadows when $x<0.005$ while anti-shadows for a sufficiently large range $x = (0.005, 0.2)$; gluon distribution of nDS nuclear PDFs shadows rather mildly at $x<0.01$ and has almost no nuclear modification for $x = (0.01, 0.4)$. Particularly there is no anti-shadowing region for gluon distribution in nDS parametrization. 
\bef
\psfig{file=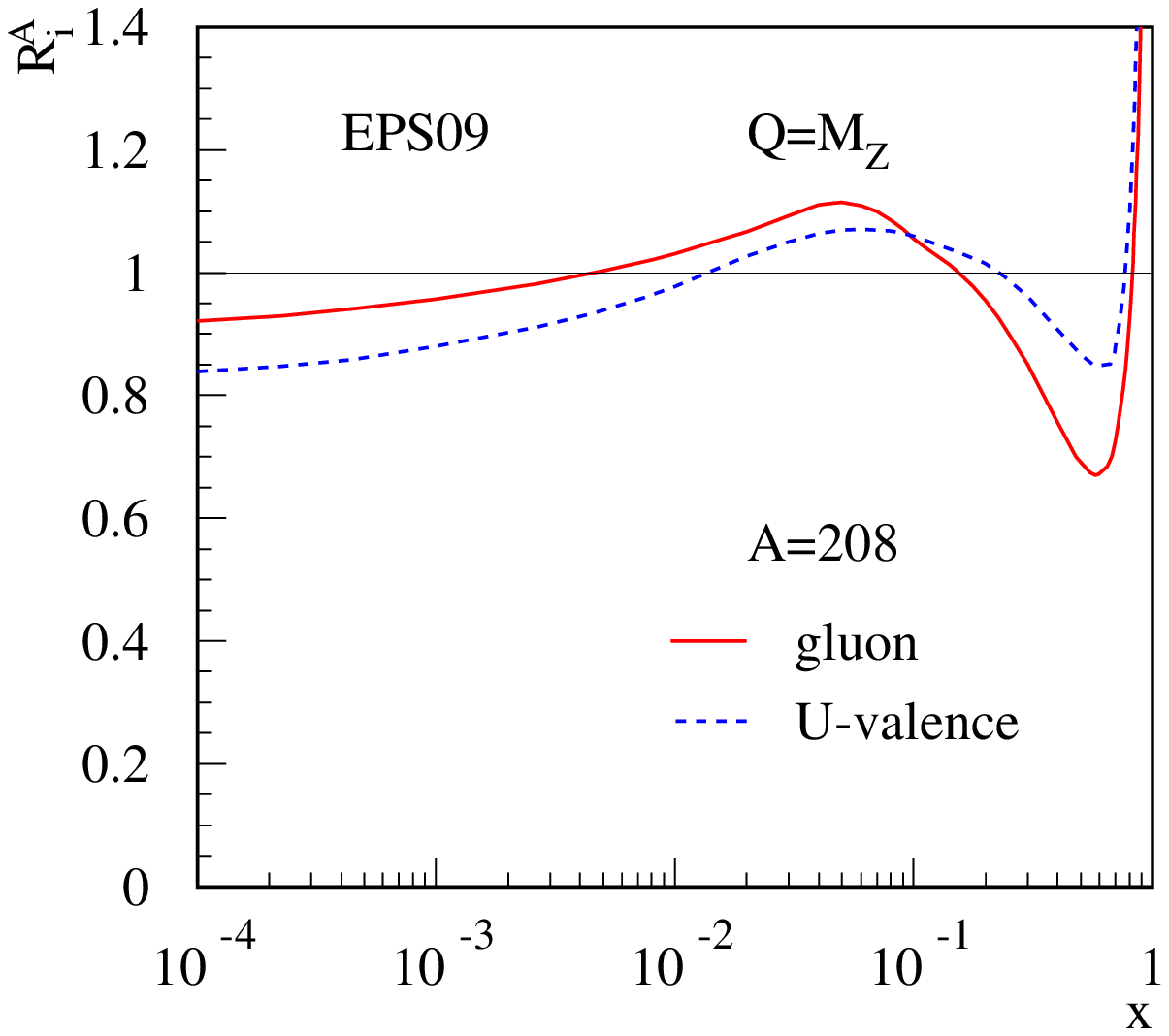, width=3in}
\hskip 0.3in
\psfig{file=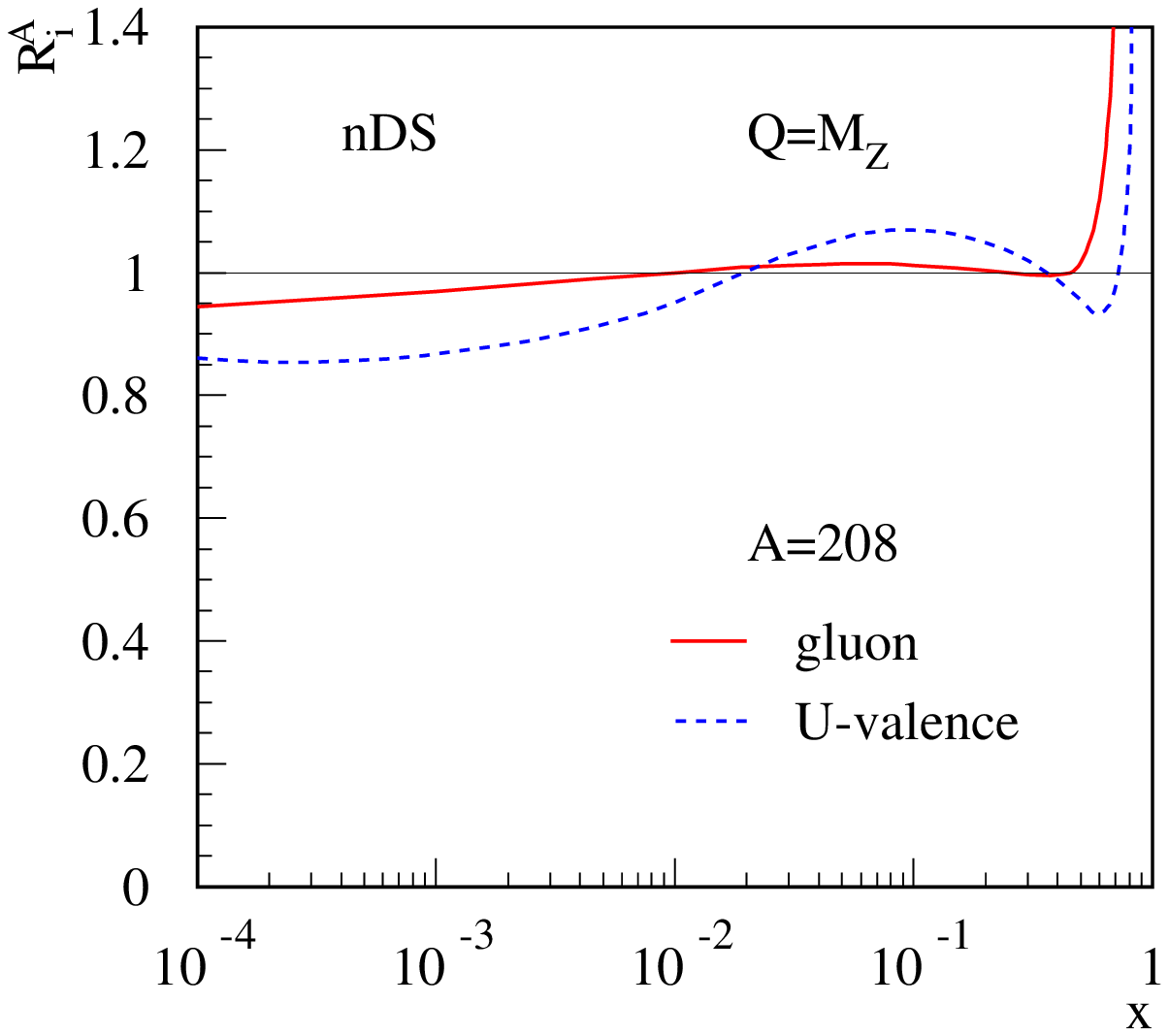, width=3in}
\caption{
\label{fig:eps09}
Ratio of nuclear PDFs over the proton PDFs at scale $Q=M_Z$: $R_{i}^A = f_{i/A}(x, Q^2)/f_{i/p}(x, Q^2)$. The red solid curve is for gluon, and the blue dashed curve is for the valence $u$-quark. Left panel is for EPS09 \cite{Eskola:2009uj} and right panel is for nDS \cite{deFlorian:2003qf}.}
\eef

Let us now take a look at the nuclear modification factor $R_{pA}$ for $Z^0$ production in p+Pb collisions at $\sqrt{s}=5$ TeV and mid-rapidity $y=0$. As indicated by the lower panel in Fig.~\ref{fig:z0} (left),  the nuclear modification factor $R_{pA}$ for $Z^0$ production in EPS09 parametrization is suppressed at low $p_T < 10$~GeV, while it is enhanced at high $p_T$, even at $p_T$ as large as 70 GeV.  The suppression at low $p_T$ was an immediate result of the shadowed EPS09 nuclear PDFs at small $x$, while the strong enhancement for such a large range of $p_T$ was a surprise.  After a careful examination of kinematics, we find that at $y=0$ the minimum parton momentum fraction $x = M_Z/\sqrt{s} \sim 0.018$, which is already in the anti-shadowing region of nuclear gluon distribution of EPS09.  That is, the clear enhancement of $R_{pA}$ in large $p_T$ region in Fig.~\ref{fig:z0} (left) is caused by the large anti-shadowing region of nuclear gluon distribution of EPS09. On the other hand, for the nDS parametrization, we find that 
the enhancement of $R_{pA}$ in the large $p_T$ disappears, as clearly seen in the lower panel in Fig.~\ref{fig:z0} (right). This is due to the absence of (or much smaller) gluon anti-shadowing of nuclear PDFs in nDS parametrization. 

In conclusion, the $R_{pA}$ of $Z^0$ production is  a direct measurement of nuclear gluon distribution since the cross section is dominated by gluon initiated subprocess when $p_T> 20$~GeV.  In particular, the measurement of nuclear modification factor $R_{pA}$ of $Z^0$ production in p+Pb collisions at the LHC provides a clean and unique test of the nuclear gluon anti-shadowing proposed in EPS09.

\section{Transverse momentum broadening of vector boson ($J/\psi$, $\Upsilon$, $W/Z^0$) production in proton-lead collisions at the LHC}  
In this section, we will study the transverse momentum broadening of vector boson ($J/\psi$, $\Upsilon$, $W/Z^0$) production in proton-lead collisions at the LHC. For the hadronic production of inclusive vector bosons, $A(p_A)+B(p_B)\to V[J/\psi,\Upsilon, W/Z^0](q)+X$ at the LHC, we define the averaged transverse momentum square of the produced vector boson as, 
\ben
\langle q_T^2\rangle(y)_{AB}
\equiv 
\int dq_T^2 \, q_T^2\,
     \frac{d\sigma_{AB\to V}}{dy\, dq_T^2}
\left/
\int dq_T^2 \, 
     \frac{d\sigma_{AB\to V}}{dy\, dq_T^2}
\right. \, .
\label{avg-qt2}
\een
We further define transverse momentum broadening in proton-lead collisions as,
\ben
\Delta\langle q_T^2\rangle_{pPb}(y)
\equiv 
\langle q_T^2\rangle(y)_{pPb}
- \langle q_T^2\rangle(y)_{pp} \, ,
\label{eq:pt-broad}
\een
i.e., the difference in the averaged transverse momentum square of vector boson production in p+Pb and p+p collisions. Transverse momentum broadening $\Delta\langle q_T^2\rangle_{pPb}$ is directly sensitive to the parton multiple scattering inside a large nucleus, as shown in Refs.~\cite{Kang:2008us,Kang:2011bp,Xing:2012ii}.

Following the derivation in Refs.~\cite{Kang:2008us,Xing:2012ii}, we have the first nonvanish contribution to
the transverse momentum broadening of heavy quarkonium production as 
\ben
\Delta\langle q_T^2\rangle_{\rm HQ}^{\rm CEM}
=\left(\frac{8\pi^2\alpha_s}{N_c^2-1}\, \lambda^2\, A^{1/3}\right) 
\frac{(C_F+C_A)\, \sigma_{q\bar{q}}+2\,C_A\, \sigma_{gg} + \Delta\sigma_{gg}}
     {\sigma_{q\bar{q}}+\sigma_{gg}} 
\label{cem-qt2}
\een
where the superscript ``CEM'' indicates that the formation of heavy quarkonium from a produced heavy quark pair is evaluated in terms of the Color Evaporation Model (CEM) \cite{CEM}, similar result was derived in NRQCD model \cite{Kang:2008us}. To derive this result, we have taken into account the initial-state multiple scattering between the incoming parton (quark or gluon) from the projectile proton and soft partons of the nuclear target, as well as the final-state rescattering between the outgoing heavy quark pair and the large nucleus. The $\sigma_{q\bar{q}}$ and $\sigma_{gg}$ in Eq.~\eqref{cem-qt2} are contributions from quark-antiquark and gluon-gluon subprocess, respectively \cite{Kang:2008us},
\ben
\sigma_{q\bar{q}}
&=&F_{Q\bar{Q}\to H}
\int_{4m_Q^2}^{4M_Q^2}dQ^2
\sum_q\int dx' \phi_{\bar{q}/p}(x')
\int dx\, \phi_{q/A}(x)\, 
\frac{d\hat{\sigma}_{q\bar{q}\to Q\bar{Q}}}{dQ^2}\, ,
\\ 
\sigma_{gg}
&=&F_{Q\bar{Q}\to H}
\int_{4m_Q^2}^{4M_Q^2}dQ^2\int dx' \phi_{g/p}(x')
\int dx\, \phi_{g/A}(x)\, 
\frac{d\hat{\sigma}_{gg\to Q\bar{Q}}}{dQ^2} \, ,
\een
where $m_Q$ is the heavy quark mass, $M_Q$ is the mass of the lowest open heavy flavor meson. $F_{Q\bar{Q}\to H}$ is a non-perturbative transition probability for a heavy quark pair $Q\bar{Q}$ transforming into a heavy quarkonium $H$ and is independent of the color and angular momentum of the heavy quark pair. $\frac{d\hat{\sigma}_{q\bar{q}\to Q\bar{Q}}}{dQ^2}$ and $\frac{d\hat{\sigma}_{gg\to Q\bar{Q}}}{dQ^2}$ represent the total cross sections for $q\bar{q}\to Q\bar{Q}$ and $gg\to Q\bar{Q}$ channels given in \cite{Kang:2008us}. We reproduce here for convenience,
\ben
\frac{d\hat{\sigma}_{q\bar{q}\to Q\bar{Q}}}{dQ^2}&=& \frac{N_c^2-1}{ N_c^2}\frac{\pi\alpha_s^2}{3Q^2} \left(1+\frac{1}{2}\gamma\right)\sqrt{1-\gamma},
\\
\frac{d\hat{\sigma}_{gg\to Q\bar{Q}}}{dQ^2} &=& \frac{1}{N_c}\frac{\pi\alpha_s^2}{Q^2} \left[
\left(1+\gamma+\frac{1}{2(N_c^2-1)}\gamma^2\right)\ln\frac{1+\sqrt{1-\gamma}}{1-\sqrt{1-\gamma}}\right.
\nnu
&&
\left.
~~~~~~~~~
-\left(\frac{5N_c^2-3}{3(N_c^2-1)}+\frac{11N_c^2-6}{6(N_c^2-1)}\gamma\right)\sqrt{1-\gamma}
\right],
\een
where $N_c=3$ is the number of colors, and $\gamma=4m_Q^2/Q^2$. 

The $\Delta\sigma_{gg}$ is a small color suppressed correction to the gluon-gluon subprocess first derived in Ref.~\cite{Xing:2012ii} \footnote{This piece was missed in the original paper Ref.~\cite{Kang:2008us}.}. It is given by
\ben
\Delta\sigma_{gg\to Q\bar Q} =  F_{Q\bar Q\to H} \int_{4m_Q^2}^{4M_Q^2} dQ^2
\phi_{g/p}(x') \int dx\, \phi_{g/A}(x)\frac{d\Delta\hat{\sigma}_{gg\to Q\bar{Q}}}{dQ^2},
\een
with $\frac{d\Delta\hat{\sigma}_{gg\to Q\bar{Q}}}{dQ^2}$ given by
\ben
\frac{d\Delta\hat{\sigma}_{gg\to Q\bar{Q}}}{dQ^2}
 =  -\frac{1}{N_c^2-1}\frac{\pi\alpha_s^2}{Q^2}
\left[\left(1+\gamma-\frac{1}{2}\gamma^2\right)
\ln\frac{1+\sqrt{1-\gamma}}{1-\sqrt{1-\gamma}}
-(1+\gamma)\sqrt{1-\gamma}\right]
\een
In the region where the gluon-gluon subprocess dominates the heavy quarkonium
production rate, $\sigma_{gg}\gg \sigma_{q\bar{q}}, \, \Delta\sigma_{gg}$, the broadening of heavy quarkonium production could be further simplified as \cite{Kang:2008us}
\ben
\Delta\langle q_T^2\rangle_{\rm HQ}^{\rm CEM}
\approx
2\, C_A
\left(\frac{8\pi^2\alpha_s}{N_c^2-1}\, \lambda^2\, A^{1/3}\right) 
\, . 
\een

In Fig.~\ref{fig:broadening}, we plot our predictions for the transverse momentum broadening of Drell-Yan type vector boson production in p+Pb collisions at the LHC at  $\sqrt{s}=5$ TeV and $y=0$ as a function of $N_{\rm coll}$, the number of collisions.  To get the dependence on $N_{\rm coll}$, effectively, we replace the $A^{1/3}$ in Eq.~(\ref{cem-qt2}) by $A^{1/3} N_{\rm coll}(b)/N_{\rm coll}(b_{\rm min.bias})$.  For the p+Pb collisions at the LHC, $N_{\rm coll}(b_{\rm min.bias}) \sim 7$ from a Glauber model calculation with nucleon-nucleon inelastic cross section $\sigma_{\rm NN}^{\rm in}=70$ mb at $\sqrt{s}=5$ TeV \cite{d'Enterria:2003qs}.  For the broadening of heavy quarkonium production, we assume that physical heavy quarkonia are formed outside the nucleus, or equivalently, we assume a maximum broadening from the final-state multiple scattering.  If the color of produced heavy quark pairs is neutralized far within the size of the lead nucleus, the rate of multiple scattering of the pairs with the nuclear medium should be reduced, so as the broadening, due to the color singlet nature of the pairs.  That is, the difference between our predictions and the measured broadening of heavy quarkonia at the LHC should provide valuable information on the color nature of produced heavy quark pairs and the time scale of color neutralization of the pairs when they transmute into physical quarkonia.

In Fig.~\ref{fig:broadening}, we also plot the broadening of $W/Z^0$ production in p+Pb collisions at the LHC by using the broadening formalism derived in Ref.~\cite{Kang:2008us}. The main difference in the transverse momentum broadening between the heavy quarkonium ($J/\psi$, $\Upsilon$) and $W/Z^0$ production is caused by the following facts: (1) $W/Z^0$ production has only initial-state multiple parton scattering, while heavy quarkonium can receive contributions from both initial-sate and final-state multiple parton scattering; and (2) the multiple scattering effect is enhanced for heavy quarkonium production because it is dominated by gluon initiated subprocesses.  Thus, the dramatic difference in broadening between the heavy quarkonium and $W/Z^0$ production in Fig.~\ref{fig:broadening} should be a signature prediction of QCD calculations.  The difference is sensitive to how a heavy quark pair transmutes itself into a physical quarkonium, and in particular, how the color of the pair is neutralized during the hadronization process, which has a direct impact on the final-state multiple scattering and its contribution to the amount of the broadening. For example, if the heavy quarkonium is created almost instantaneously as a singlet state, then the final-state multiple scattering effect will be much smaller due to small color dipole moment of a tightly bound color singlet heavy quark pair.  In this case, the size of the broadening will be significantly reduced (by about a factor of two according to our formalism Eq.~\eqref{cem-qt2}).
\bef
\psfig{file=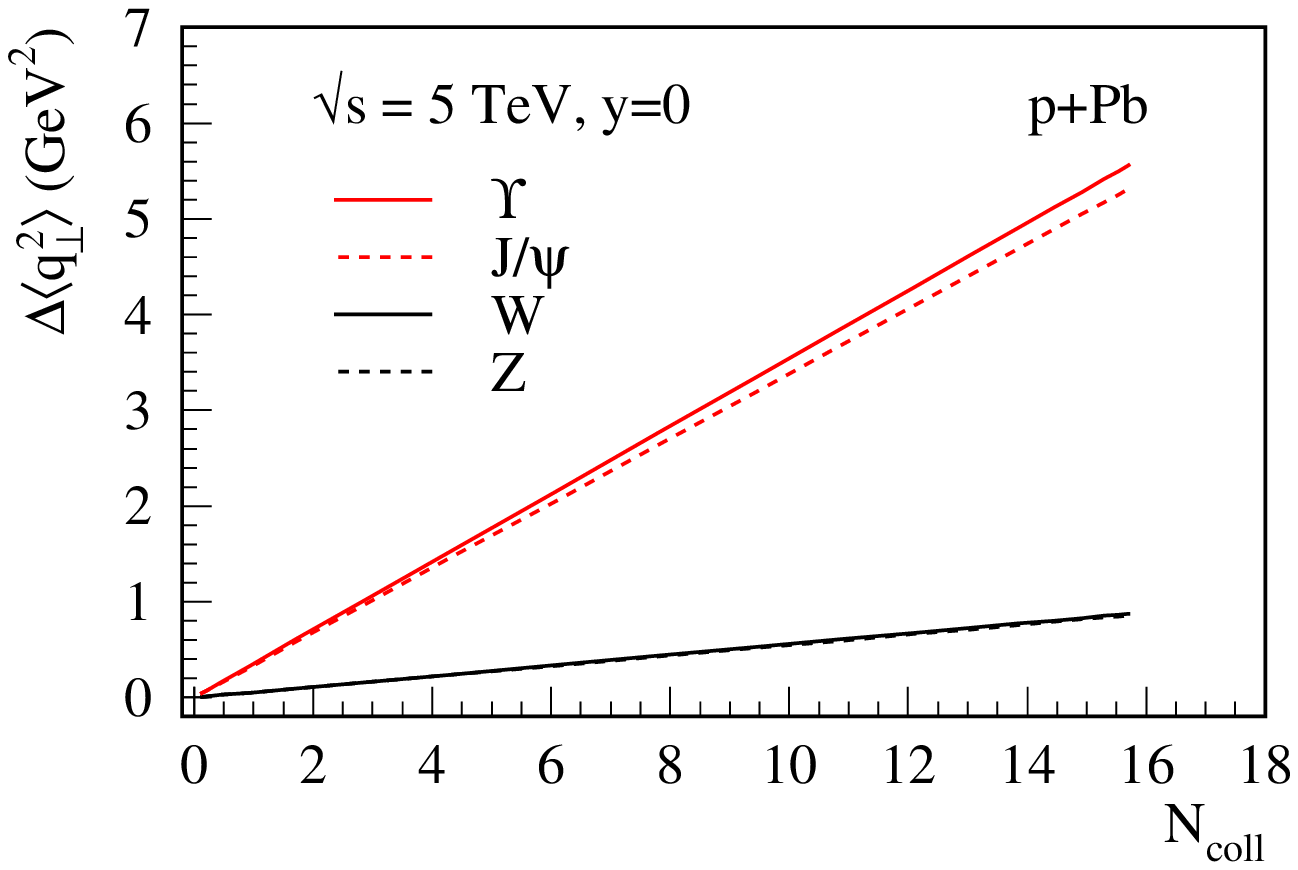, width=3.4in}
\caption{Transverse momentum broadening of vector boson production in p+Pb collision is plotted as a function of number of collisions $N_{\rm coll}$, we choose LHC kinematics $\sqrt{s}=5$ TeV and mid-rapidity $y=0$. Red solid is quarkonium $\Upsilon$, red dashed is $J/\psi$, black solid is $W$, and black dashed is $Z^0$ boson.}
\label{fig:broadening}
\eef

\section{Conclusions}
In this paper, within the CSS resummation formalism, we first studied the nuclear modification of transverse momentum distribution of $Z^0$ production in p+Pb collisions at LHC energy $\sqrt{s}=5$ TeV and mid-rapidity $y=0$.  We found that the production rate of $Z^0$ in this kinematic region is dominated by the gluon initiated subprocesses when transverse momentum $p_T > 20$ GeV.  Consequently, the nuclear modification factor $R_{pA}$ is an excellent observable to measure the nuclear modification of gluon distribution in a large nucleus, which is practically unknown.   Very different nuclear gluon distributions were extracted from various QCD global analyses of nuclear PDFs.  In this paper, we presented predictions for nuclear modification factor $R_{pA}$ of $Z^0$ production as a function of $p_T$ in p+Pb collisions, which will soon be confronted by new experimental data in the upcoming proton-lead run at the LHC.  In particular, we found that with EPS09, nuclear modification factor is suppressed at low $p_T < 10$ GeV, while it is enhanced at high $p_T$, even at $p_T$ as large as 70 GeV. This clear enhancement of $R_{pA}$ in large $p_T$ region is caused by the large anti-shadowing region of nuclear gluon distribution of EPS09. The large enhancement at large $p_T$ actually  disappears if we use another parametrization of nuclear PDFs (such as nDS) which has much smaller (or no) gluon anti-shadowing in the nuclear gluon distribution. Therefore, the measurement of nuclear modification factor $R_{pA}$ of $Z^0$ production in p+Pb collisions at the LHC provides a clean and unique test of the nuclear gluon anti-shadowing proposed in EPS09. 

We also made predictions for the transverse momentum broadening of vector boson ($J/\psi$, $\Upsilon$, $W/Z^0$) production in proton-lead collisions at $\sqrt{s}=5$ TeV at the LHC. We found that the transverse momentum broadening for heavy quarkonium production is much larger than those for $W/Z^0$ production, due to the difference in color flow of the dominated partonic subprocesses for the production of heavy quarkonia and heavy $W/Z^0$ bosons, and the additional final-state parton multiple scattering for heavy quarkonium production. The dramatic difference in broadening between the heavy quarkonium and $W/Z^0$ production is a signature prediction of QCD calculations of parton multiple scattering in a better-controlled cold nuclear medium. 

\section*{Acknowledgments}
This work was supported in part by the US Department of Energy, Office of Science, under Contract No.~DE-AC52-06NA25396 (Z.K.) and DE-AC02-98CH10886 (J.Q.). \\

Note added. When this paper was being prepared, a preprint \cite{Guzey:2012jp} has just appeared which also studied the nuclear modification of $Z^0$ boson production at the LHC. Ref.~\cite{Guzey:2012jp} focused primarily on the small-$x$ or shadowing region, while we study the anti-shadowing region.

\end{document}